\documentclass[journal,comsoc]{IEEEtran}
\usepackage{cite}
\usepackage{amsmath,amssymb,amsfonts}
\usepackage{algorithmic}
\usepackage{graphicx}
\usepackage{textcomp}
\usepackage{xcolor}
\usepackage{comment}
\usepackage{listings}
\usepackage{fancyvrb}

\usepackage{blkarray}
\usepackage{amsmath}
\usepackage{dsfont}
\usepackage{amsthm}
\usepackage{lipsum}
\usepackage{graphicx}

\usepackage[T1]{fontenc}% optional T1 font encoding
\usepackage{fancyhdr}

\ifCLASSINFOpdf
\else
\fi
\usepackage{amsmath}
\usepackage[cmintegrals]{newtxmath}
\begin{document}
%\chead{text}{\emph{This work has been submitted to the IEEE for possible publication. Copyright may be transferred without notice, after which this version may no longer be accessible.}}
\title{\begin{scriptsize}\textcolor{blue}{\emph{This work has been submitted to the IEEE for possible publication. Copyright may be transferred without notice, after which this version may no longer be accessible.}} \end{scriptsize}\\
A Proactive Connection Setup Mechanism for Large Quantum Networks}

%\author{Author 1,
%        Author 2,
%        Author 3,
%       and Author 4% <-this % stops a space
%\thanks{M. Shell was with the Department
%of Electrical and Computer Engineering, Georgia Institute of Technology, Atlanta,

\author{Dibakar~Das,~\IEEEmembership{Member,~IEEE,}
        Shiva~Kumar~Malapaka, %~\IEEEmembership{Fellow,~OSA,}
        Jyotsna~Bapat,~\IEEEmembership{Member,~IEEE,}
        and~Debabrata~Das,~\IEEEmembership{Senior~Member,~IEEE}% <-this % stops a space

%\thanks{M. Shell was with the Department}
%of Electrical and Computer Engineering, Georgia Institute of Technology, Atlanta,
}

\maketitle

\begin{abstract}
Quantum networks use quantum mechanics properties of entanglement and teleportation to transfer data from one node to another. Hence, it is necessary to have an efficient mechanism to distribute entanglement among quantum network nodes. Most of research on entanglement distribution apply current state of network and do not consider using historical data. This paper presents a novel way to quicken connection setup between two nodes using historical data and proactively distribute entanglement in quantum network. Results show, with quantum network size increase, the proposed approach improves success rate of connection establishments.
\end{abstract}


% \fi
\IEEEpeerreviewmaketitle

\section{Introduction}\label{section_introduction}

\begin{figure}[ht]
\centering
\includegraphics[width=3in]{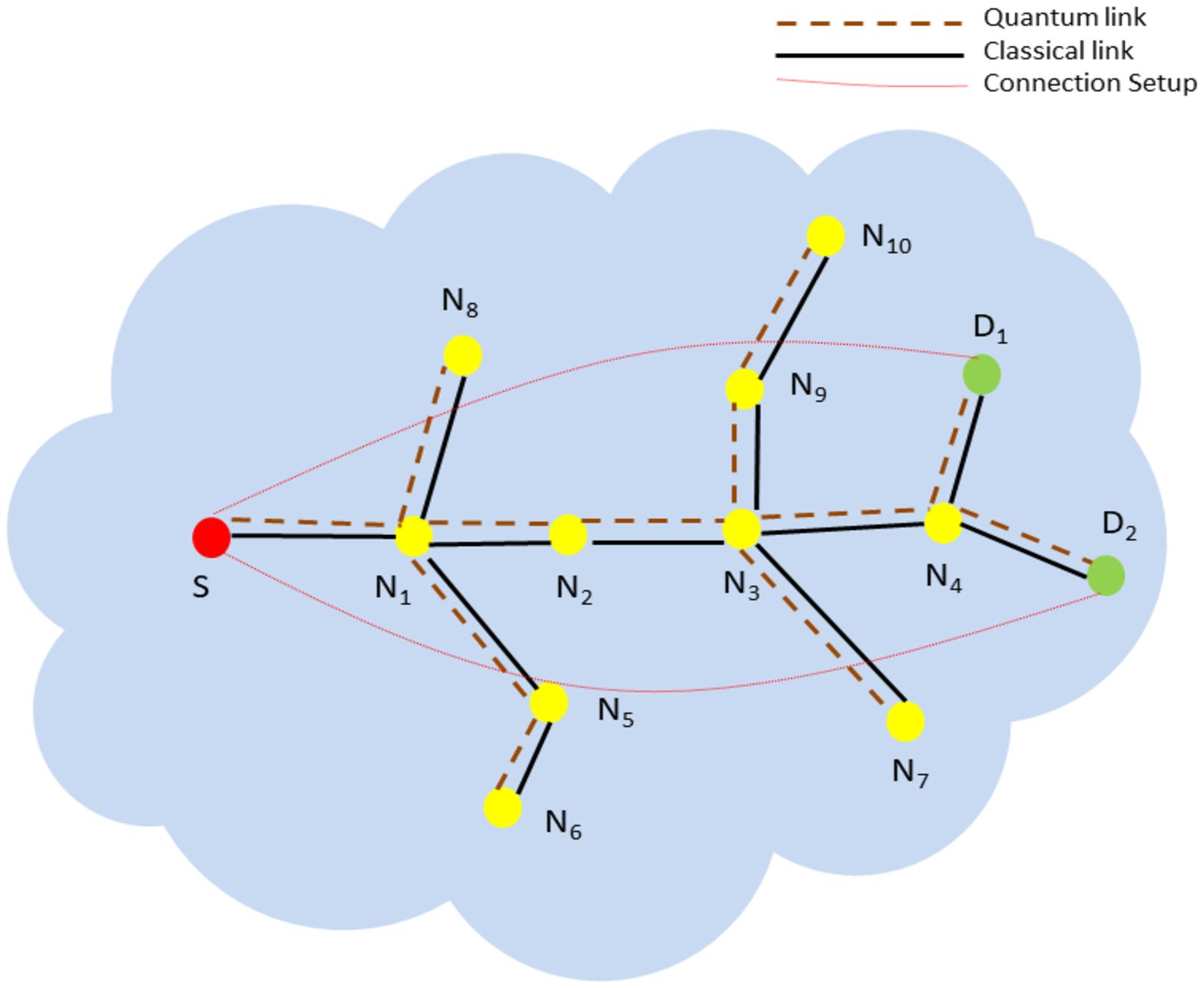}
\caption{A typical quantum network}
\label{plot_QI_Network_Example_inkscape}
\end{figure}

\IEEEPARstart{Q}{uantum} networks (QNet) is a major research area in recent years. These networks exploit quantum mechanics principles, such as, quantum entanglement (QEnt) and quantum teleportation (Qtel), to communicate between distant nodes. Due to lossy nature of quantum channel over larger distances (loss of fidelity), one way to set up a connection between two remote quantum nodes (QNode) is usage of multiple hops at short distances, e.g., quantum repeaters (QRep). Apart from each QNode having quantum and classical interfaces to its neighbours, for large QNet, nodes are expected to generate and transmit QEnt pairs, and store QEnt pairs in quantum memory (QMem) in future \cite{cite_qi_link_layer}.
QTel needs a classical channel to send the results of quantum measurement (QMeas) on the entangled qubits to its peer, which the latter uses to infer the data sent by the former.
Each QNode has quantum (brown dashed lines) and classical (black solid lines) interfaces as show in Fig. \ref{plot_QI_Network_Example_inkscape}. End-to-end quantum connection are shown with red dotted lines.

Quantum heralded entanglement (QHEnt) has been favoured for initial deployment of QNet. QHEnt \cite{cite_quantum_heralded_ent} works better with lesser quantum resources, such as, less qubits and no QMem. QHEnt has a timing constraints that both the photons from the two nodes need to reach the heralding station at the same time. However, for large deployment with QHEnt alone may not scale up when quality of service (QoS) comes into play across remote nodes. 
Hence, quantum entanglement swapping (QEntSwap) would be necessary for future QNet, when large number of entangled qubits could be generated and stored in QMem to tranfer information \cite{cite_qi_link_layer}. QNet using quantum error correction codes \cite{cite_ent_without_mem_quantum_error_correction} and quantum network coding \cite{cite_net_coding_vs_swapping} avoiding entanglement generation and swapping altogether have also been proposed. However, they require much more higher node capabilities and may not be attempted in the near future \cite{cite_qi_link_layer}, and hence, this topic is not considered any further in this work.

%Literature Survey
Significant research has happened on various aspects of QNet. A link layer protocol for QNet has been described in \cite{cite_qi_link_layer}.
%A vision for quantum internet has also been explored.\cite{cite_qi_vision}
Also, researchers have suggested quantum network coding for entanglement distribution over QEntSwap \cite{cite_net_coding_vs_swapping}. Novel protocols for QReps along with generation of entanglement simultaneously with multiple user pairs with enhanced performance have been developed \cite{cite_rout_ent_for_qi}. 
A mechanism to opportunistically distribute QEnt in a network using a local minimum of a cost function with QMem error as input, and predicting evolving entanglement fidelities has been proposed \cite{cite_opportunistic_ent_dist_qi}. A QEntSwap protocol for heterogeneous quantum devices (using particle and wave qubits) to set up hybrid entanglement between distance QNodes has also been proposed\cite{cite_quantum_swap_hetero_encoding} which will help implementing the proposal in this work a possibility.
Internet Engineering Task Force (IETF) has also initiated studies on quantum internet providing
high level overview of connection setup \cite{cite_ietf_connect_setup_qi} without going into detailed implementable specifications at this point of time.

As evident from the above discussion, most of the recent works in QNet deal with current state of quantum parameters.
This paper exploits QEntSwap to create proactive links based on history of  number of QEnt qubits [history count (HC)] used along each quantum link to speed up connection setup between remote nodes. Though, QEntSwap requires more entanglement qubits, it can avoid the need of QHEnt at every hop. 
QEnt qubits can be allocated to the links (based on HCs) where they are mostly likely to get used. QEnt between two nodes can be set up using QHEnt (wherever possible/necessary) or through QEntSwap.
Simulation, results show that setting up QEnts proactively among some of the intermediate QNodes lead to higher success in setting up end-to-end connection between two remote nodes.
To the best of the knowledge of the authors, none of the previous work have explored this novel proposal.

Assumptions made in this proposal are as follows. Each QNode has a classical and a quantum interface. Each QNode has capability to generate QEnt (QHEnt or QEntSwap), can save entangled qubits in stable QMem till they are used for data transfer, and can perform QEntSwap. The HCs (not the qubits themselves) used along each link between QNodes are saved in classical memory. All control procedures, such as, connection setup are performed over classical interface and data transfer happens using QTel over quantum interface.
\section{System Model}\label{section_system_model}
Lets consider the QNet in Fig. \ref{plot_QI_Network_Example_inkscape}. The black solid lines represent quantum links. Each edge of QNode $N_i$ has HC of $q_i$. Higher HC indicates more data was transferred through those links (using QTel). Lets consider the node $N_1$. $N_1$ is directly connected with QNodes $N_2$, $N_5$ and $N_8$ with HCs $q_2$, $q_5$ and $q_8$ respectively. $N_1$ calculates the mean $\mu_{N_1}$ as follows.
\begin{equation}\label{eqn_mean}
\mu_{N_1} = \frac{q_2 + q_5 + q_8}{3}
\end{equation}
Then, $N_1$ evaluates the squared differences between  $\mu_{N_1}$ and each of $q_2$, $q_5$ and $q_8$ as follows.
\begin{equation}\label{eqn_mean_diff_sqd}
\Delta_{\mu_{N_1}^{(i)}} = (\mu_{N_1} - q_i)^2
\end{equation}
$i = 2, 5, 8$

$N_1$ takes the minimum of the $\Delta_{\mu_{N_1}^{(i)}}$ and selects $N_2$ to set up a proactive QEnt (brown dashed lines in Fig. \ref{plot_QI_Network_Example_inkscape}), assuming $\Delta_{\mu_{N_1}^{(2)}}$ is the minimum. Similarly, $N_3$ selects $N_4$ to set up QEnt. This mechanism is applied to avoid giving more weight to the maximum HC of the neighbour links and minimize use of precious entangled qubits.

Further, $N_{10}$ can only select $N_{9}$, whereas $N_{9}$ can choose between $N_{3}$ and $N_{10}$ by using the same mechanism as already explained. $N_2$ has to select $N_{3}$. In this way, each QNode in the QNet can set up a QEnt with one of its neighbours (provided HC is not zero). If $N_3$ selects $N_4$, then in the next step $N_1$ can set up QEnt with $N_3$ using QEntSwap. Once this step is done, $N_1$ can set up QEnt with $N_4$ again with QEntSwap. This way the QEnt can build up in the overall network and make the QNet ready to higher layer connection setup between two QNodes. These proactive QEnts based on HC speed up the connection setup process when the QNet is large. 

Once a higher layer request for a connection setup between $S$ and $D_1$ is received, then only two QEnts have to be set up. $S$ has to set up QEnt with $N_1$ and $N_4$ has to set up with $D_1$. Then, applying QEntSwap first between $S$ and $N_4$ via $N_1$ followed by the same between $S$ and $D_1$ via $N_4$ the complete connection can be set up between $S$ and $D_1$ (red dotted lines in \ref{plot_QI_Network_Example_inkscape}). If there is a subsequent request to set up a connection between $S$ and $D_2$, then only one QEnt is required between $N_4$ and $D_2$. A QEntSwap between $S$ and $D_2$ via $N_4$ sets up the new connection. Thus, the previous set of QEnts and QEntSwap performed at each stage while helps in reducing the number of such steps while setting up future connections.

Connection setup request between any two QNodes in QNet is sent by the higher layers in the protocol stack. If the source and destination QNodes are already entangled then the connection setup can be done without any further QEnts or QEntSwaps.
If the two QNodes are not entangled, then the source QNode searches for QNodes which have entanglement with it. The entangled neighbour QNode (of the source), which in turn has maximum count of QEnts with its neighours, is chosen as the next hop. %Searching for the next hop will require a protocol to be designed at each QNode.
Source QNode will query its entangled neighbours to report their respective count of QNodes they are entangled with. Based on this report, the source QNode will choose the entangled neighbour which has the maximum count as the next hop. Once the next hop is decided, the connection setup request is forward to the selected QNode.

It is possible that during this process of forwarding the connection setup request from one hop to another, the request may get into a circular loop. Hence, it is necessary that the connection setup keeps track of QNodes that it has already visited (which can be piggy backed on the connection setup message). Note that when QEnts are set up between QNodes proactively this list of visited nodes will be short (compared to IETF approach \cite{cite_ietf_connect_setup_qi}). Once, a QNode detects the next hop is one in the list already visited, it randomly selects another QNode with which it is entangled. If it has no entanglements with any other QNode, it chooses any of its next hop neighbours to set up QEnt and forwards the message. Apart from these steps, to break the loop there are two decrementing counters, one for selecting a random entangled QNode and the other for selecting a neighbour at random. When the first counter reaches 0 (unable to break the cycle) the second counter is started. When the second counter also reaches 0 a connection \emph{failure} is declared. After that a new retry is made for the same connection. Each trial to set up a connection leads to setting up a few more QEnts, which benefits later retries. All these steps continue till the next hop is the destination node as requested by the higher layers in the connection setup message and the connection setup is a \emph{success}. Once the connection is setup between two QNodes data transfer can happen using QTel. HCs can be updated when data transfer happens using QTel.

\section{Results and Discussion}\label{section_results}
\begin{figure}[ht]
\centering
\includegraphics[width=3in]{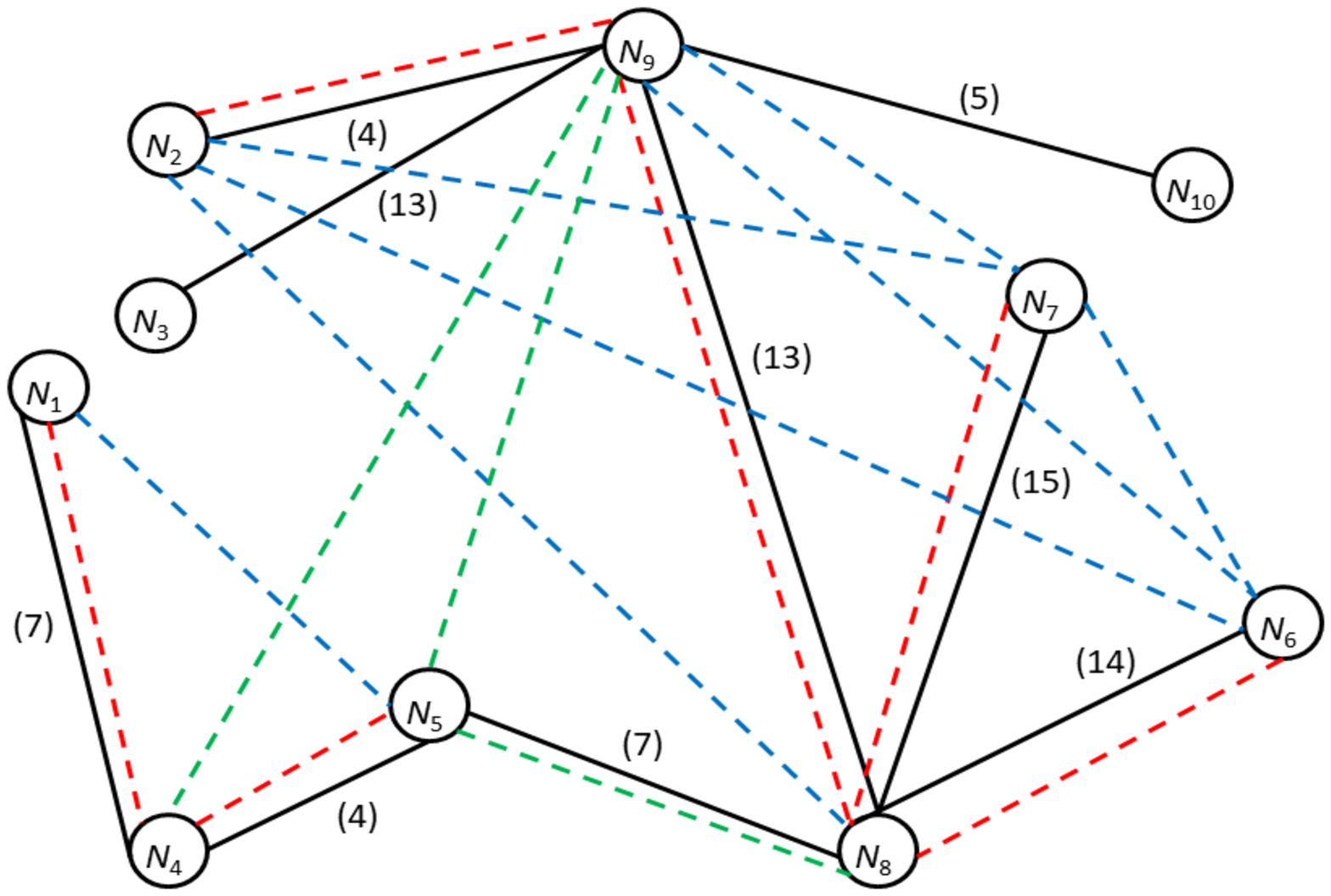}
\caption{Simulated QNet with proactively generated QEnts - solid black lines are physical links and dashed lines are QEnts}
\label{fig_single_iter_example_qnet_inkscape}
\end{figure}

Lets consider the QNet in Fig. \ref{fig_single_iter_example_qnet_inkscape} generated randomly. The QNet has 10 nodes. The black edges are the physical links (both classical and quantum). The number at each edge  shows the HCs, i.e., $q_i$ values. Using (\ref{eqn_mean}) and (\ref{eqn_mean_diff_sqd}), $\Delta_{\mu_{N_1}^{(i)}}$ for each edge is calculated. For example, $N_8$ has 4 options $N_5$, $N_6$, $N_7$ and $N_9$ to set up QEnt. The $\Delta_{\mu_{N_1}^{(i)}}$ for edges to $N_5$, $N_6$, $N_7$ and $N_9$ are 27.5625, 3.0625, 7.5625, 0.5625 respectively. Hence, $N_9$ is selected by $N_8$ to set up QEnt which has the minimum $\Delta_{\mu_{N_1}^{(i)}}$. Nodes which have only one link (in black) $N_6$ and $N_7$ set up QEnt along those edges. However, $N_5$ has two options $N_4$ and $N_8$ with same $\Delta_{\mu_{N_1}^{(i)}}$ value of 2.25. Hence, $N_5$ chooses randomly $N_4$ to set up QEnt. Similarly, $\Delta_{\mu_{N_1}^{(i)}}$ are calculated for other nodes. Finally, the list of QEnts set up based on HCs are shown in red dashed lines, namely, $\{N_4, N_1\}$, $\{N_8, N_9\}$, $\{N_5, N_4\}$, $\{N_2, N_9\}$, $\{N_7, N_8\}$, $\{N_6, N_8\}$. Based on this initial set of QEnts, QEntSwap can be applied to set up further QEnts, namely, $\{N_1, N_5\}$, $\{N_2, N_8\}$, $\{N_6, N_7\}$, $\{N_6, N_9\}$, $\{N_7, N_9\}$  and subsequently $\{N_6, N_2\}$, $\{N_7, N_2\}$ (all shown in blue dash lines in Fig. \ref{fig_single_iter_example_qnet_inkscape}), as explained in section \ref{section_system_model}.

Lets consider a connection setup request from $N_4$ (source) and $N_9$ (target). $N_4$ neither has a direct physical link with $N_9$ nor entangled. So, $N_4$ tries to find QEnt nodes with maximum entanglement count. Incidently, both $N_1$ and $N_5$ has same number of entanglements, i.e., 2 ($N_1$ has with $N_4$ and $N_5$, and $N_5$ has between $N_1$ and $N_4$, only blue and red dash lines in Fig. \ref{fig_single_iter_example_qnet_inkscape} should be considered, green dash lines form later). So, $N_4$ randomly chooses  $N_1$ as next hop. However, $N_1$ also does not have direct physical link with $N_9$ nor a QEnt link. Hence, $N_1$ too tries the same way with $N_4$ and $N_5$ and selects the former randomly. Now, $N_4$ realizes a cycle and selects $N_5$. $N_5$ neither have a direct link nor a QEnt with $N_9$, and hence it has the only option to select $N_8$ as next hop and set up QEnt (green dashed line). Now, $N_5$ can set up QEnt with $N_9$ with QEntSwap (green dashed line) via $N_8$. Subsequently, $N_4$ can set up QEnt with $N_9$ using QEntSwap via $N_5$. Thus, a connection between $N_4$ and $N_9$ is setup. Now, data transfer can happen between $N_4$ and $N_9$ using QTel.

\begin{figure}[ht]
\centering
\includegraphics[width=2.5in]{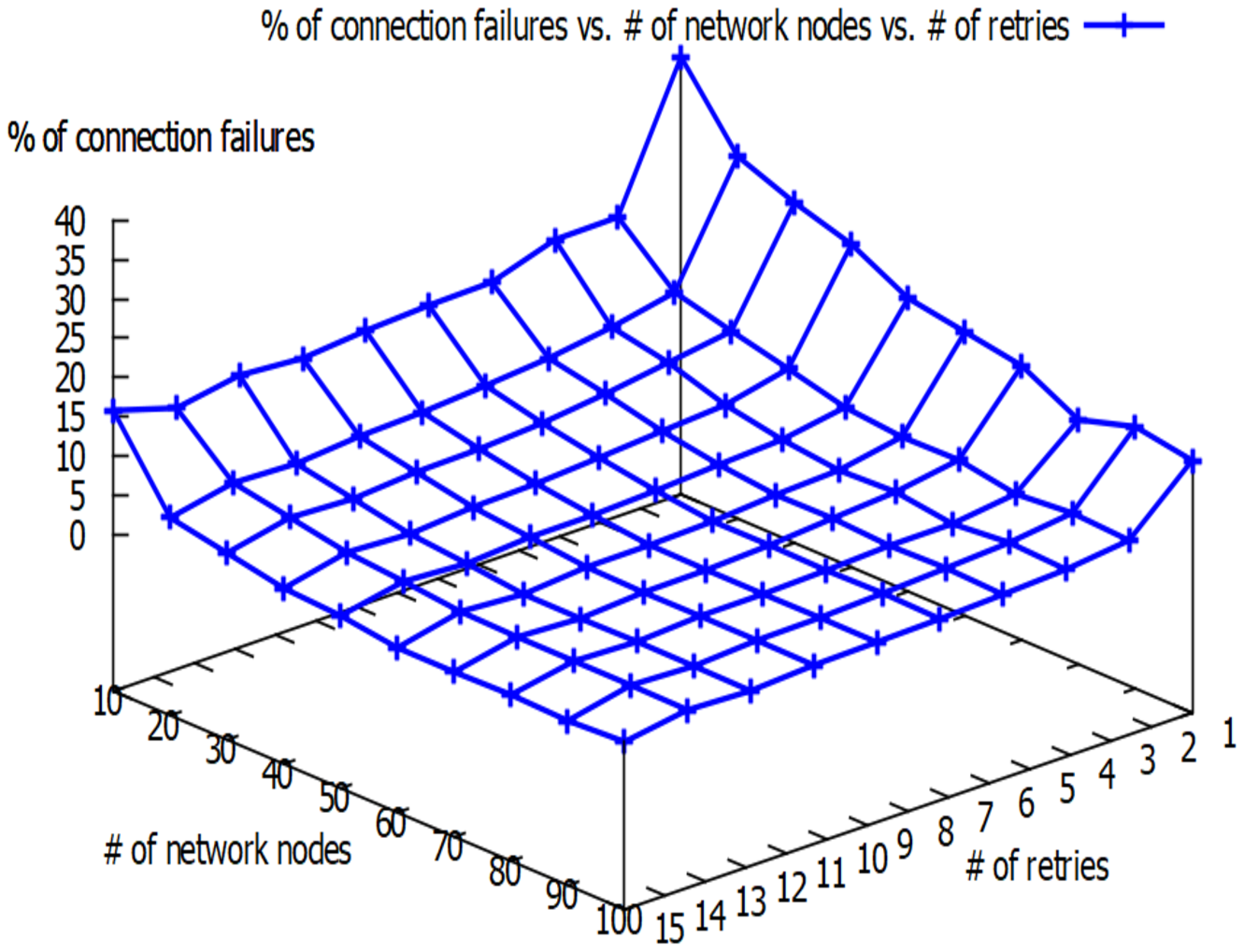}
\caption{Single connection setup behaviour with increase in \# of retries and \# of nodes}
\label{fig_plot_connection_setup_1c_10to100nodes_inkscape}
\end{figure}

Fig. \ref{fig_plot_connection_setup_1c_10to100nodes_inkscape} shows the behaviour of first connection setup in a QNet using the proposed algorithm. Along \emph{x}-axis, number of connection retries are shown, whereas \emph{y}-axis depicts number of nodes in the QNet. Connection setup failures is shown along \emph{z}-axis. A couple of interesting conclusion can be drawn from Fig. \ref{fig_plot_connection_setup_1c_10to100nodes_inkscape}. Firstly, the connection setup failures drop with increase in number of retries for the first connection. This is because subsequent retries trigger further QEnts between nodes which bring significant decrease in failures of future connection setup in the QNet. Secondly, connection setup failures drop even further with combined increase in number of retries and number of nodes in the QNet due to similar reason mention above. Thirdly, not much variation in connection failure rate is observed with number of nodes greater than 40 and retries greater than 5. If a connection is retried infinitely eventually it will succeed but it will impact QoS.
These conclusions can help in deciding on network size and also number of retries while setting up connections between remote (likely unentangled) nodes.

\begin{figure}[ht]
\centering
\includegraphics[width=2.5in]{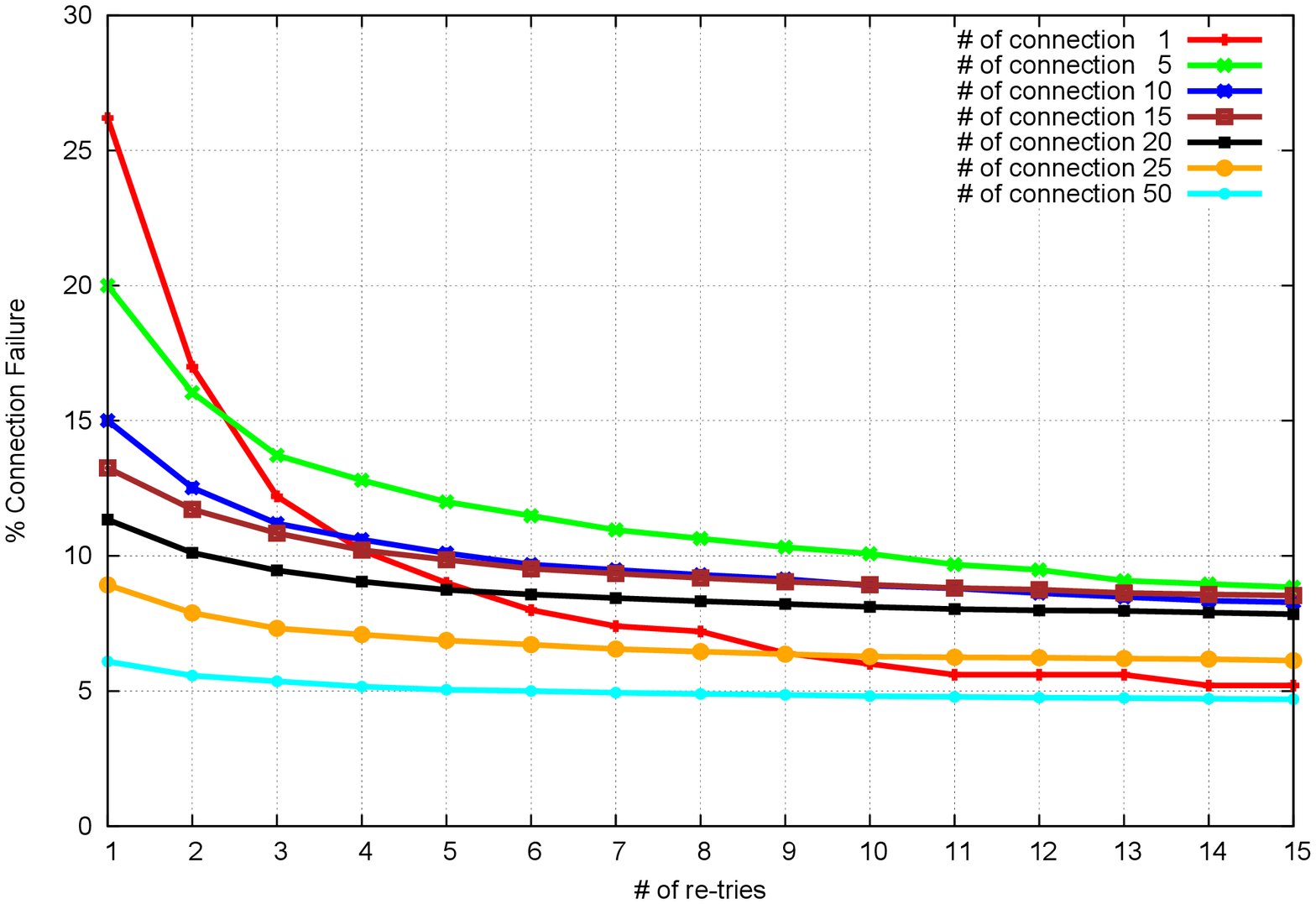}
\caption{Multiple connection setup (1-50) behaviour with increase in \# of retries}
\label{fig_plot_connection_setup_1to50_20nodes_inkscape}
\end{figure}

\begin{figure}[ht]
\centering
\includegraphics[width=2.5in]{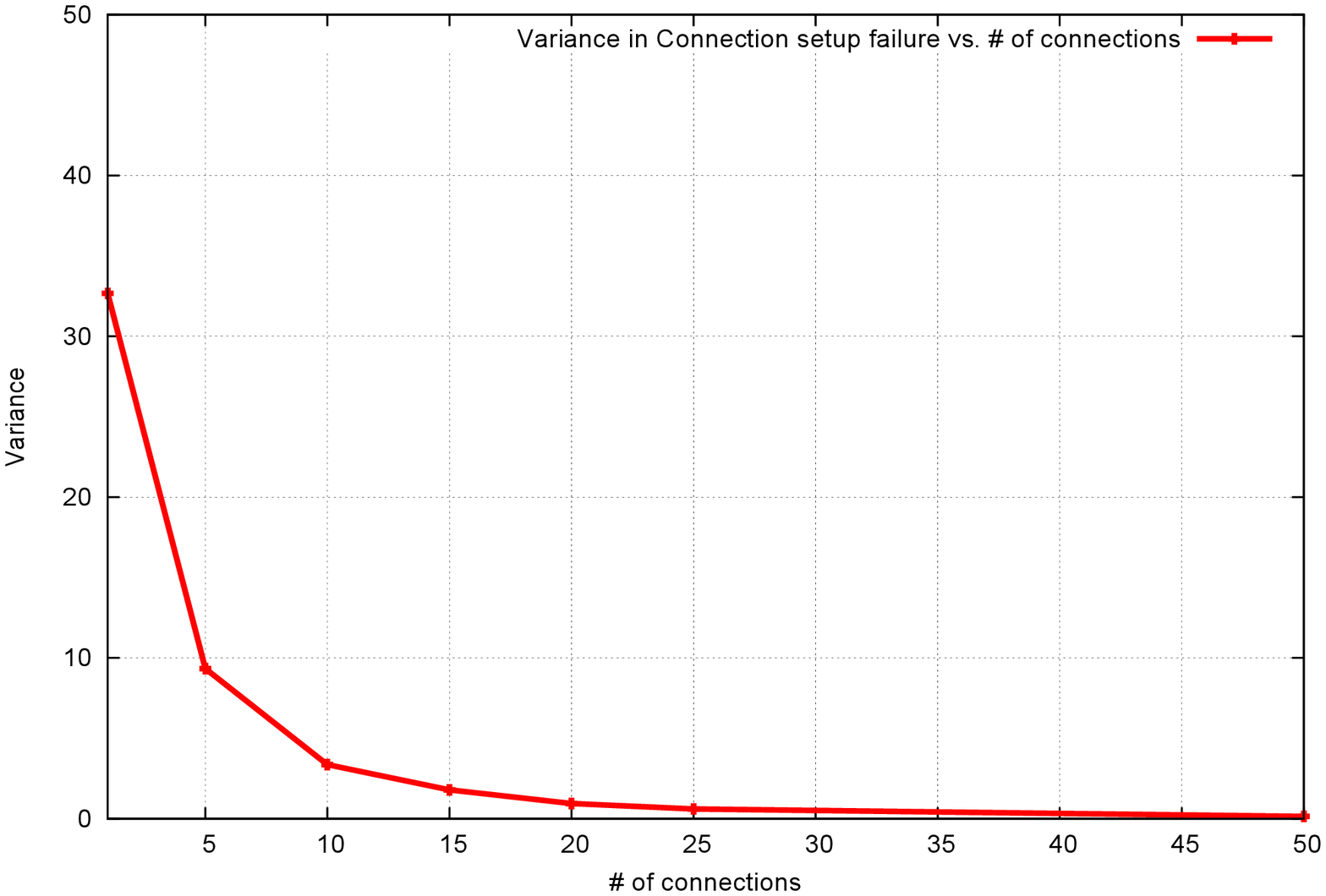}
\caption{Variance in connection setup failures with increase in \# of connections}
\label{fig_plot_connection_setup_1to50_20nodes_variance_inkscape}
\end{figure}

Fig. \ref{fig_plot_connection_setup_1to50_20nodes_inkscape} shows the behaviour when number of connections are increased from 1 to 50 in a fixed size QNet of 20 nodes. Connection failure rate and number of retries are respectively depicted along \emph{y} and \emph{x} axes. Following facts are observed. For first connection setup in a network, it starts poorly with high failure rate but improves significantly with increase in retries. Increase in number of connections further from 5 to 50 shows consistent decrease in failures. The reason behind this is that each connection setup attempt helps in setting up QEnt between pairs of nodes which leads to successful setup of later connections. Increase in retries leads to further reduction in failures. Interestingly, the variation in error due to retries for a given number of connections also reduces due to reasons mentioned above already. This is also evident from the variance plotted against number of connections in Fig \ref{fig_plot_connection_setup_1to50_20nodes_variance_inkscape}. In fact, for 50 connection setups the performance is better than setting up a single connection. Also, for retries $\leq$ 4, setting up 10 or more connections are more efficient than setting up one connection. Beyond 10 retries the variations in connection failures are still less for any number of connections. This essentially means that retry counter can be less when large number of connection setups are expected.

\begin{figure}[ht]
\centering
\includegraphics[width=2.5in]{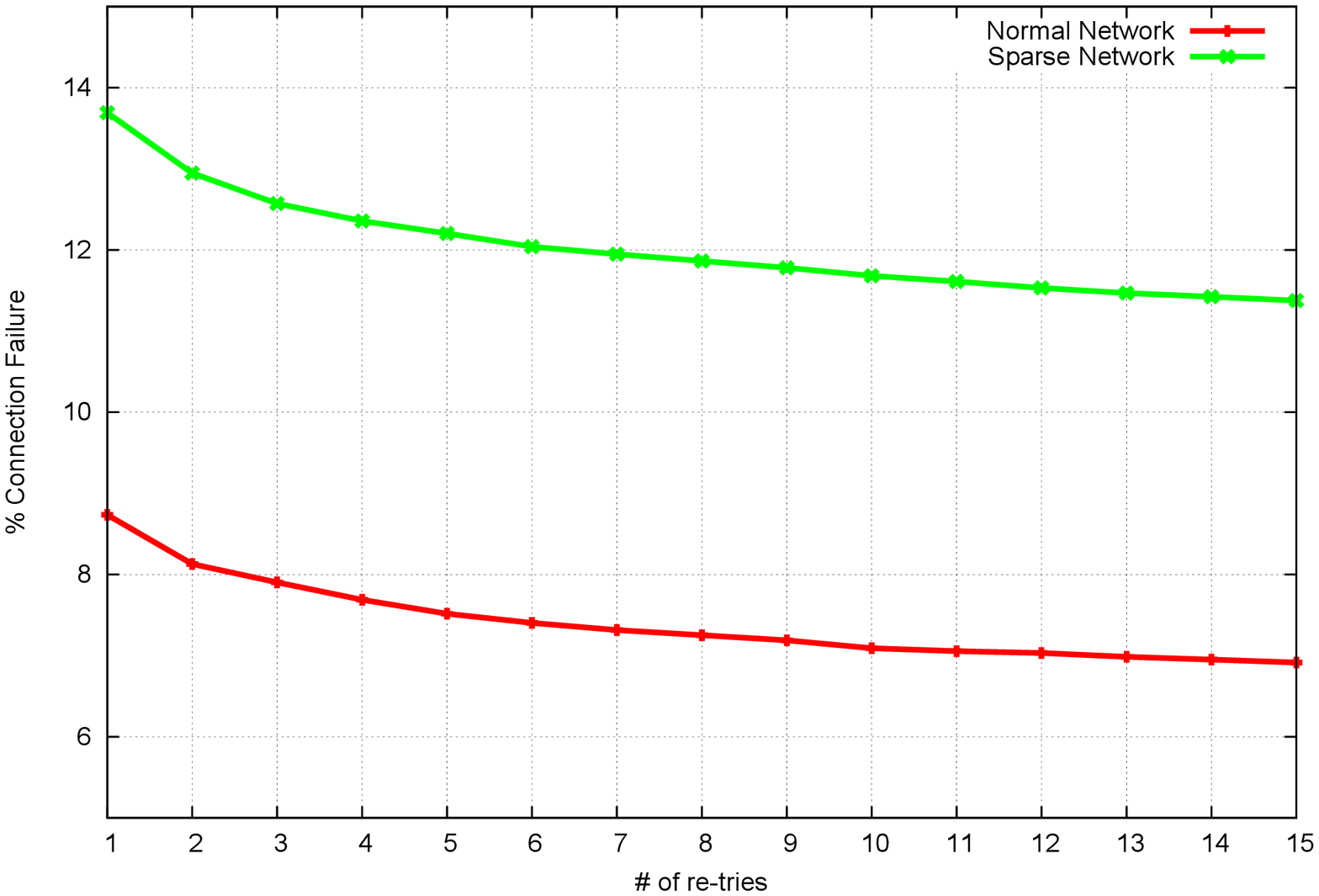}
\caption{Comparison of connection setup failures in a normal vs. sparse quantum network with increase in \# of retries}
\label{fig_plot_connection_setup_50connection_100nodes_normal_sparse_inkscape}
\end{figure}

%\subsection{Connection setup in normal versus sparse network}
Fig. \ref{fig_plot_connection_setup_50connection_100nodes_normal_sparse_inkscape} shows the behaviour when the network is made sparse as compared to above results. Labels along \emph{x} and \emph{y} axes remain the same as before. The average number of neighbours of a node in the QNet is decreased by half to make the network spare. It can be observed that increase in sparsity leads to higher connection failures.

IETF approach \cite{cite_ietf_connect_setup_qi} suggests creating QEnts only after receiving a higher layer request to set up a connection between two nodes. The QEnts are set up as the QNet tries to route the connection setup request message through it. If the proposed idea in this paper is not applied then setting up a connection between $N_1$ and $N_{10}$ (in Fig. \ref{fig_single_iter_example_qnet_inkscape} along the physical links with black lines) would require QEnts set up along $N_1$, $N_4$, $N_5$, $N_8$, $N_9$, $N_{10}$ followed by QEntSwaps. Along all these nodes path information has to be carried to set up the route to the target node and then acknowledged back to the source node. This approach also slows down the connection setup process. However, with the proposed idea of proactive QEnts, $N_1$ to $N_{10}$ would require QEntSwaps between $N_1$ and $N_9$ (in Fig. \ref{fig_single_iter_example_qnet_inkscape} with all the necessary QEnts set up with red, blue and green dash lines), QEnt between $N_9$ and $N_{10}$ and then finally a QEntSwap between $N_1$ and $N_{10}$. The path information that has to be carried in the connection setup message is also shorter compared to the IETF method \cite{cite_ietf_connect_setup_qi}.

\section{Conclusion and Future Work}\label{section_conclusion}
Data transfer over a QNet is possible using QTel. QEnt is required for QTel. Hence, it is necessary to distribute QEnt across nodes of a large QNet efficiently to set up a connection between two likely unentangled node. Though several research works have happened in this direction, none of the previous proposals consider use of HC of QEnt qubits. This work proposed a novel proactive way to distribute QEnt in the network guided by HC to help speed up connection setup between two remote nodes. Simulation results show that using the proposed approach significant reduction in connection setup failure can be achieved. This proposed idea can contribute to efficient design of large QNet.

The proposed work considers only HC at a node for setting up QEnts with its neighbours having physical links.
Also, while selecting the next hop during proactive QEnt distribution, a combination of maximum value of HC among neighbours and averages can be used.
\section*{Acknowledgment}
The authors would like to thank Tejas Networks for supporting this research work.

\ifCLASSOPTIONcaptionsoff
  \newpage
\fi



%\bibliographystyle{IEEEtran}
%\begin{thebibliography}{1}


%\end{thebibliography}

\bibliographystyle{IEEEtran}
\bibliography{IEEEabrv,QI-Connection-Setup}

%\end{IEEEbiography}


%\newpage



%\vfill


\end{document}